\documentclass[conference]{IEEEtran}
\IEEEoverridecommandlockouts
\usepackage{cite}
\usepackage{amsmath,amssymb,amsfonts}
\usepackage{algorithmic}
\usepackage{graphicx}
\usepackage{textcomp}
\usepackage{xcolor}
\usepackage{dblfloatfix}
\usepackage[french]{babel}
\usepackage[T1]{fontenc}
\usepackage{hyperref}
\def\BibTeX{{\rm B\kern-.05em{\sc i\kern-.025em b}\kern-.08em
    T\kern-.1667em\lower.7ex\hbox{E}\kern-.125emX}}
\begin{document}

\title{MULTISS: un protocole de stockage confidentiel à long terme sur plusieurs réseaux QKD\thanks{This work has been conducted within the framework of the French government financial support managed by the Agence Nationale de la Recherche (ANR), within its Investments for the Future programme, under  the Stratégie Nationale Quantique through the project of the PEPR-quantum QComtestbeds (ANR 22-PETQ-0011) and with fundings from the EUROPE HORIZON-FPA project QSNP and the EUROPE DIGITAL project FranceQCI.}}

\author{\IEEEauthorblockN{Thomas Prévost}
\textit{Université Côte d'Azur}\\
\IEEEauthorblockA{\textit{I3S - CNRS} \\
Sophia-Antipolis, France \\
thomas.prevost@univ-cotedazur.fr}
\and
\IEEEauthorblockN{Olivier Alibart}
\textit{Université Côte d'Azur}\\
\IEEEauthorblockA{\textit{InPhyNi - CNRS} \\
Nice, France \\
olivier.alibart@univ-cotedazur.fr}
\and
\IEEEauthorblockN{Anne Marin}
\textit{VeriQloud}\\
\IEEEauthorblockA{marin@veriqloud.fr}
\and
\IEEEauthorblockN{Marc Kaplan}
\textit{VeriQloud}\\
\IEEEauthorblockA{kaplan@veriqloud.fr}
}

\maketitle

\begin{abstract}
Cet article présente MULTISS, un nouveau protocole pour le stockage à long terme, distribué sur plusieurs réseaux de distribution quantique de clé (Quantum Key Distribution, QKD). Ce protocole est une extension de LINCOS, un protocole de stockage sécurisé qui utilise le partage de secret de Shamir pour le stockage de secret sur un seul réseau QKD. Notre protocole utilise le partage de secret hiérarchique pour distribuer un secret entre plusieurs réseaux QKD tout en garantissant la sécurité parfaite. Notre protocole permet en outre de mettre à jour les partages sans avoir à reconstruire tout le secret. Nous prouvons également que MULTISS est strictement plus sécurisé que LINCOS, qui demeure vulnérable lorsque son réseau QKD est compromis.
\end{abstract}


\section{Introduction}
Certaines informations sont classées confidentielles sur le long terme, au moins plusieurs décennies. C'est le cas par exemple de certains secrets industriels, de secrets d'\'{E}tat ou de données médicales. Ces secrets ont donc besoin d'une garantie de confidentialité et de disponibilité à long terme. Les primitives cryptographiques actuelles sont cependant inadaptées pour de tels usages. En effet, la technologie et la cryptanalyse évoluent rapidement. On sait par exemple que les primitives de chiffrement à clé publique utilisées majoritairement aujourd'hui, c'est-à-dire RSA ou ECC, seront vulnérables aux attaques par ordinateur quantique~\cite{bhatia2020efficient,wohlwend2016elliptic}. De même, la cryptographie \og~post-quantique~\fg \ sera susceptible d'être vulnérable à la cryptanalyse dans un futur plus ou moins éloigné~\cite{kaluderovic2022attacks}. On peut donc légitimement craindre qu'un acteur malveillant écoute les communications chiffrées, dans le but de les déchiffrer lorsque la cryptanalyse dont il disposera le permettra. Ce type d'attaque est appelé \og~Harvest now, decrypt later~\fg~\cite{paul2022transition}.

Claude Shannon a proposé en 1949 le principe de \og~sécurité parfaite~\fg \ (ou inconditionnelle)~\cite{shannon1949communication} (Information Theoretic Secrecy, \emph{ITS}). Cela signifie que la confidentialité est assurée peu importe la puissance de calcul de l'attaquant. En d'autres termes, \og~il est aussi difficile pour un attaquant de trouver la clé que de découvrir le message par hasard~\fg. La confidentialité à long terme ne saurait être garantie que par un protocole qui garantit la sécurité parfaite tout au long de son exécution.

Si la sécurité parfaite est généralement proposée par un \og~masque jetable~\fg \ (One Time Pad), c'est à dire une clé parfaitement aléatoire non réutilisée de la même taille que le message, la distribution quantique de clé~\cite{bennett2014quantum} est une méthode originale de transmission garantissant ce même niveau de sécurité. Ce mécanisme ne repose plus sur la limite de puissance de calcul de l'attaquant, mais sur la possibilité de détecter celui-ci à la volée lorsqu'il écoute le message. Puisque l'attaquant n'est même pas en mesure d'écouter les communications, une attaque de type \og~Harvest now, decrypt later~\fg \ n'est pas envisageable.

Le protocole LINCOS~\cite{braun2017lincos} permet de conserver un secret sur un réseau de distribution quantique de clé. Il utilise pour cela le partage de secret de Shamir~\cite{shamir1979share}, afin de répartir le secret sur les différents n\oe{}uds du réseau QKD. Le partage de secret de Shamir garantissant lui-même la sécurité parfaite~\cite{corniaux2014entropy}, le protocole LINCOS offre le même niveau de confidentialité tant que l'attaquant n'a pas compromis un nombre de n\oe{}uds équivalent au seuil de déchiffrement.

Cependant, la transmission quantique de clé sur laquelle repose le protocole LINCOS est limitée par une distance géographique maximale, de l'ordre de quelques centaines de kilomètres. Les implémentations de LINCOS, en Europe~\cite{openqkdltss} comme en Asie~\cite{japanqcloud}, limitent la portée du réseau à une surface métropolitaine. De plus, les différents n\oe{}uds d'un réseau QKD partagent généralement le même sous-réseau informatique. Ainsi, des pirates informatiques qui prendraient le contrôle du réseau, ou bien une entité étatique qui pourrait saisir légalement le matériel, serait en mesure de reconstruire le secret.

Nous proposons donc une extension de LINCOS, nommée MULTISS~\cite{prevost2024mutliss}, permettant d'étendre le stockage de secret sur plusieurs réseaux de distribution quantique distants. Notre protocole conserve les propriétés de sécurité parfaite de LINCOS, tout en garantissant la confidentialité du secret contre un attaquant en mesure de prendre le contrôle de l'ensemble d'un des réseaux de QKD.

\section{La distribution quantique de clé}

En cryptographie classique, la transmission de secrets entre deux entités distantes repose généralement sur la combinaison du chiffrement à clé publique et du chiffrement à clé secrète, le chiffrement à clé publique permettant d'échanger la clé de chiffrement symétrique qui sera utilisée pour chiffrer effectivement le secret. La sécurité de ce type de protocoles est dite \og~calculatoire et sémantique~\fg, c'est à dire qu'elle repose sur la difficulté, pour un attaquant aux ressources de calcul bornées, de retrouver la clé privée à partir de la clé publique ou bien de retrouver le secret à partir du chiffré, dans un délai raisonnable. Les participants sont contraints de faire une hypothèse sur la puissance de calcul de l'attaquant qui, nécessairement, augmente avec le temps.

La distribution quantique de clé fonctionne tout à fait différemment. La sécurité est assurée par un principe fondamental de la physique quantique, le \emph{théorème de non-clonage}~\cite{zygelman2018no}. Le théorème stipule qu'il est impossible de cloner parfaitement un qubit arbitraire (la représentation de base de l'information quantique) sans modifier son état.

Les protocoles de QKD fonctionnent ainsi soit par la transmission directe de qubits entre les participants, comme c'est le cas pour le protocole BB84~\cite{BB84}, soit avec une source intermédiaire qui envoie deux qubits intriqués vers les deux participants, comme pour le protocole E91~\cite{ekert1991quantum}. Lorsqu'un adversaire intercepte un qubit pour lire son état, alors il modifie nécessairement ledit état (comme stipulé par le théorème de non-clonage). Les participants sont alors en mesure de détecter cette modification et peuvent interrompre l'échange. En général, les bits échangés par la QKD vont servir de masque jetable pour chiffrer les données~: $C = S \oplus K$, avec $S$ le message secret, $K$ les bits de clé échangés par la QKD, et $C$ le chiffré.

Demeure néanmoins la problématique d'authentifier les participants honnêtes, afin d'empêcher un attaquant de réaliser une attaque dite de \og~l'Homme du milieu~\fg \ (Man-in-the-Middle, MitM). Cette authentification est généralement réalisée grâce à la cryptographie classique. On utilise soit un échantillon d'une clé échangée antérieurement, soit un mécanisme d'authentification à clé publique. Ces mécanismes d'authentification seraient bien entendu vulnérables contre un attaquant qui disposerait d'une puissance de calcul illimitée \emph{au moment de l'échange}. En fait, nous définissons notre modèle d'attaquant avec une puissance de calcul limitée et connue, mais qui pourrait évoluer indéfiniment par la suite. On parle alors de \og~sécurité éternelle~\fg \ (everlasting security). Pour simplifier, dans la suite de cet article, nous qualifierons de \og~liens à sécurité parfaite~\fg, ou \og~liens ITS~\fg, les liaisons QKD entre deux n\oe{}uds.

En général, la représentation du qubit utilisée pour la distribution quantique de clé est l'état quantique d'un photon (par exemple sa polarisation). Le photon est transporté au sein d'une \og~fibre noire~\fg, protégée contre les interférences extérieures. Cependant, les pertes de la fibre optique limitent la portée maximale de la QKD à quelques centaines de kilomètres au maximum~\cite{Pelet}, puisque le théorème de non-clonage interdit l'usage d'un répéteur.

\section{Le partage de secret} \label{secret_sharing}

\subsection{Partage de secret direct}

Le partage de secret est une primitive découverte simultanément par Adi Shamir~\cite{shamir1979share} et George Blakley~\cite{blakley1979safeguarding} en 1979. Nous utilisons ici la primitive proposée par Shamir. Elle permet à une personne, le \emph{dealer}, de distribuer un secret entre $n$ participants, chaque fraction du secret s'appelant un \og~partage~\fg. Le dealer définit un seuil $k$ de participants qui devraient mettre leurs partages en commun afin de reconstruire le secret initial.

Dans le partage de secret de Shamir, on définit $S \in \mathbb{N}$ le secret initial, $n \in \mathbb{N}^*$ le nombre de participants et $k \in \mathbb{N}^*$ le seuil minimal de déchiffrement, défini par le dealer. N'importe quel sous ensemble de $k$ participants doit être en mesure de retrouver le secret en mettant leurs partages en commun, tandis que s'il n'y a que $k - 1$ participants, ces derniers ne disposent d'aucune information sur le secret.

Le dealer commence par choisir un grand nombre premier $q$, tel que $q > S$. Par la suite, toutes les opérations sont effectuées dans le corps fini $\mathbb{F}_q$. Le dealer génère un polynôme aléatoire $P \in \mathbb{F}_q[X]$ de degré $deg(P) = k - 1$, tel que $P(0) = S$, le secret initial. Il distribue ensuite les évaluations du polynôme $P(1), P(2), \dots, P(n)$ aux $n$ participants, qui ont chacun connaissance du degré du polynôme. Si $k$ participants parmi $n$ mettent leurs partages en commun, ils sont en mesure de reconstruire le polynôme $P$ par interpolation lagrangienne~\cite{waring1779vii}, et donc de retrouver le secret $S = P(0)$.

\subsection{Partage de secret hiérarchique}

On peut étendre le concept de partage de secret pour y ajouter une notion de hiérarchisation des participants~\cite{tassa2004hierarchical}. Imaginons par exemple un directeur de banque qui souhaiterait un minimum de 3 employés pour ouvrir le coffre. Le partage de secret hiérarchique lui permettrait d'exiger au minimum 3 employés, \textbf{dont} au moins un responsable d'agence parmi eux.

De même que pour le partage de secret de Shamir, le dealer génère un polynôme aléatoire $P \in \mathbb{F}_q[X]$ de degré $deg(P) = k - 1$, tel que $P(0) = S$. Il calcule ensuite le polynôme dérivé de $P$ par rapport à $X$, $P'$. Il distribue alors à chacun des managers les évaluations du polynôme primitif $P(1), P(2), \dots, P(m)$, puis aux employés les évaluations du polynôme dérivé $P'(1), P'(2), \dots, P'(n)$, avec $m$ le nombre de managers et $n$ le nombre d'employés.

Ainsi, $k$ employés \textbf{dont} $1$ manager sont en mesure de reconstruire le polynôme $P$ par interpolation de Birkhoff~\cite{birkhoff1906general}, et donc de retrouver le secret $S = P(0)$.

\section{Le protocole LINCOS}

\begin{figure}
\centering
\includegraphics[width=1\linewidth]{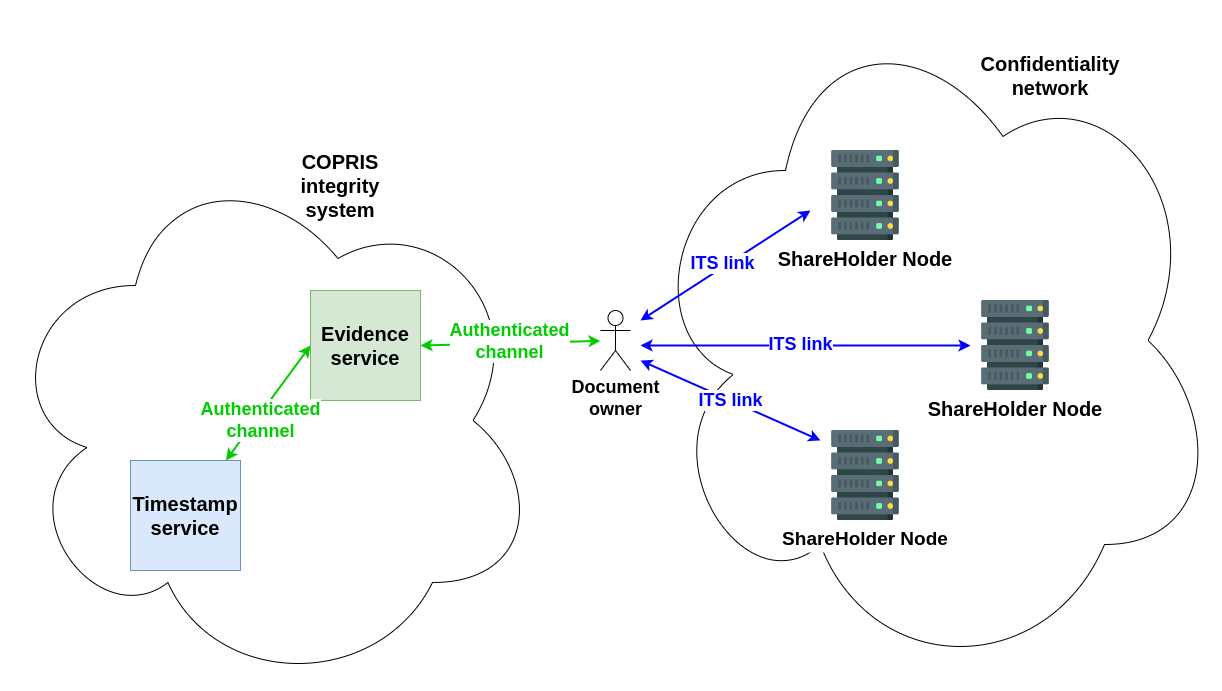}
\caption{Le protocole LINCOS, avec à gauche COPRIS, en charge de l'intégrité et de l'authentification, et à droite le stockage du document par partage de Shamir entre les différents n\oe{}uds du réseau QKD.}
\label{fig:lincos_scheme}
\end{figure}

Le protocole LINCOS~\cite{braun2017lincos}, sur lequel est basé MULTISS, est composé de deux procédures distinctes, comme décrit Fig.~\ref{fig:lincos_scheme}:

\begin{itemize}
    \item COPRIS, le protocole d'intégrité et d'authenticité~;
    \item le protocole de confidentialité à long terme.
\end{itemize}

C'est ce dernier protocole que nous entendons étendre avec MULTISS, le protocole COPRIS restant inchangé.

\subsection{COPRIS, pour l'intégrité et l'authenticité}

Afin de garantir l'intégrité et l'authenticité d'un document, son propriétaire peut utiliser COPRIS, afin de prouver que le document a existé à un instant $t$, tout en gardant ce document secret. Pour cela, le propriétaire envoie un \emph{engagement} à l' \og~Evidence service~\fg. Ce dernier fait une requête d'horodatage courant au \og~Timestamp service~\fg. A partir de l'horodatage reçu, l'evidence service créé un enregistrement.

Puisque le propriétaire refait régulièrement des requêtes d'engagement, il est en mesure de prouver qu'un document $S$ existait bien à un instant $t$.

\subsection{Le protocole de confidentialité à long terme} \label{lincos_confidential}

Dans le protocole LINCOS, le propriétaire du document dispose d'un lien ITS avec chacun des n\oe{}uds du réseau QKD. En pratique, il existe un lien quantique (une fibre noire en général) entre le propriétaire du document et les n\oe{}uds. Le propriétaire et le n\oe{}ud échangent une clé aléatoire grâce à la distribution quantique de clé, puis utilisent cette même clé comme un masque jetable pour chiffrer leurs communications. C'est donc un lien de confidentialité parfaite, compatible avec la confidentialité à long terme.

Afin de stocker son document $S$ au sein du réseau de QKD, le propriétaire choisit un seuil de déchiffrement $k$, en fonction de ses besoins de disponibilité et de la confiance qu'il accorde au réseau. Il génère alors des partages du document $S$ à partir de la primitive de Shamir, avec un seuil de déchiffrement de $k$, tel qu'expliqué en section~\ref{secret_sharing}. Il envoie alors les partages $P(1), P(2), \dots, P(n)$ aux différents n\oe{}uds du réseau. Il faudra alors mettre en commun au moins $k$ partages pour retrouver le document $S$.

Le propriétaire du document doit en outre régulièrement mettre à jour les partages sur les différents n\oe{}uds, afin de rendre improbable l'exploitation de potentielles fuites de données successives, c'est à dire si un attaquant était en mesure de prendre le contrôle de plusieurs n\oe{}uds à des moments distincts. Afin de ne pas reconstruire le document secret à chaque mise à jour des partages, le propriétaire génère un nouveau polynôme aléatoire $Q \in \mathbb{F}_q[X]$ de degré $k - 1$, tel que $Q(0) = 0$. Il envoie alors les évaluations $Q(1), Q(2), \dots$ aux différents n\oe{}uds, qui vont additionner les valeurs reçues à leurs anciennes valeurs (toujours dans le corps fini $\mathbb{F}_q$). Ainsi, il sera toujours possible de reconstruire un polynôme $R = P + Q$, avec $R(0) = P(0) + Q(0) = S + 0 = S$, à partir d'au moins $k$ partages.

\section{Hypothèses et modèle d'adversaire}

\subsection{Hypothèses de sécurité}

Le protocole MULTISS permet d'étendre le protocole LINCOS sur plusieurs réseaux de QKD, de sorte que la compromission d'un réseau entier n'affecte pas la confidentialité du secret. Dans le protocole LINCOS, le propriétaire du fichier est relié par un \og~lien ITS~\fg \ avec le réseau de QKD. Ce lien ITS consiste en une liaison de distribution quantique de clé vers les n\oe{}uds du réseau, bien que nous pourrions par exemple imaginer un PDG d'entreprise qui amènerait en personne son document secret dans un mallette sécurisée. Puisque les différents sous-réseaux du protocole MULTISS sont censés se trouver très éloignés géographiquement, alors nous faisons l'hypothèse que le propriétaire du document est relié par un lien ITS à un seul de ces sous-réseaux. Les autres étant hors de portée des liens quantiques, ils sont reliés au propriétaire du document par un lien n'autorisant que la cryptographie classique. On appellera par la suite \og~réseau mère~\fg \ le sous-réseau relié au client par un lien ITS, et \og~réseaux filles~\fg \ les autres sous-réseaux. À noter cependant que cette dénomination varie en fonction de l'emplacement géographique des propriétaires~: le sous-réseau métropolitain parisien sera le réseau mère pour un client français, mais un réseau fille pour un client japonais.

De plus, nous formulons l'hypothèse raisonnable que les mécanismes d'authentification utilisés ne sont pas vulnérables au moment où les messages sont échangés. Ainsi, on suppose que l'authentification est parfaite, et qu'il est donc impossible pour un adversaire de réaliser une attaque de l'Homme du milieu.

\subsection{Compromission d'un réseau entier par un adversaire} \label{comp_adv}

De par la limitation géographique de la QKD, les \og~réseaux ITS~\fg \ se déploient généralement sur une surface métropolitaine. Ils sont en outre souvent gérés par la même entité administrative, par exemple un laboratoire de recherche ou une entreprise, et donc connectés au même sous-réseau IP. Il est ainsi possible qu'un attaquant qui a réussi à prendre le contrôle d'un des n\oe{}uds puisse aisément se déplacer dans le réseau interne. La limitation géographique de l'étendue de ces réseaux de QKD pose aussi un problème juridictionnel. Un état aujourd'hui démocratique pourrait ne pas le rester à l'avenir, et un dictateur pourrait ordonner la saisie \og~légale~\fg \ de tous les n\oe{}uds du réseau, découvrant ainsi le secret.

MULTISS permet de se protéger contre un adversaire qui pourrait potentiellement prendre le contrôle d'un sous-réseau ITS entier.

\subsection{Harvest now, decrypt later} \label{hndl_adv}

Puisque le propriétaire du document n'est relié par un lien ITS qu'à un seul des sous-réseaux, le réseau mère, alors la communication avec les autres sous-réseaux filles se fait nécessairement par cryptographie classique. Le protocole MULTISS garantit la confidentialité persistante contre un adversaire \og~Harvest now, decrypt later~\fg, qui serait en mesure d'écouter puis de déchiffrer ultérieurement les communications passées par des liens classiques.

\subsection{Incompatibilité des deux modèles d'adversaires}

Les deux modèles d'adversaire décrits ci-avant sont incompatibles entre eux. Un adversaire qui aurait écouté toutes les communications sur les liens classiques qu'il serait ensuite en mesure de déchiffrer (sect.~\ref{hndl_adv}), puis qui prendrait le contrôle du réseau ITS mère (sect.~\ref{comp_adv}) serait forcément en mesure de retrouver le document secret initial, puisqu'il disposerait de l'ensemble de l'information existante. MULTISS ne permet de se protéger que d'un seul de ces deux adversaires à la fois.

Cela ne devrait cependant pas poser un problème pour la garantie de confidentialité à long terme. En effet, la compromission du sous-réseau ITS mère (par voie \og~légale~\fg \ ou piratage informatique) a peu de chances de demeurer discrète sur le long terme pour le propriétaire du document. Celui-ci aura alors le temps de prendre les dispositions nécessaires pour invalider le contenu du document. De même, si le chiffrement utilisé sur les canaux classique venait à devenir vulnérable, le propriétaire finirait aussi par en être informé.

\section{Description du protocole MULTISS}

MULTISS utilise deux niveaux de partage de secret. De plus, à la différence de LINCOS, le propriétaire ne doit plus définir un mais trois seuils distincts:

\begin{itemize}
    \item $t_{nodes}$ le nombre minium de n\oe{}uds à contrôler pour retrouver le secret, tous sous-réseaux confondus~;
    \item $t_{networks}$ le nombre minimum de sous-réseaux à compromettre pour retrouver le secret. Un sous-réseau est considéré comme \og~compromis~\fg \ dès lors qu'un seul de ses n\oe{}uds a été compromis~;
    \item $t_{fail}$ le nombre minimum de n\oe{}uds à arrêter pour que le secret ne soit plus accessible.
\end{itemize}

En fonction de ces seuils, le propriétaire définira les degrés des différents polynômes, comme nous le verrons plus tard. À noter qu'il est possible de ne choisir qu'un certain nombre de combinaisons de seuils, en fonction de la topologie du réseau.

Le propriétaire du document commence par générer un polynôme aléatoire $P \in \mathbb{F}_q[X]$, tel que $P(0) = S$, le document secret. Puis le propriétaire génère $l$ polynômes $Q_i \in \mathbb{F}_q[X]$ tels que $Q_0(0) = P(1)$, et $Q_i(0) = P'(i) \ \forall i \in [\![ 1, l-1 ]\!]$, avec $l$ nombre de sous-réseaux, et $P'$ dérivé de $P$. Finalement, le propriétaire répartit les évaluations $Q_0(1), Q_0(2), \dots$ sur les n\oe{}uds du réseau mère, et chaque évaluation $Q_i(1), Q_i(2), \dots$ sur les n\oe{}uds de chaque sous-réseau fille $i$.

Pour reconstruire le secret, le propriétaire aura besoin des partages $Q_j(0), j \in [\![ 0, l-1 ]\!]$ d'au moins $deg(P) - 1$ sous-réseaux \textbf{dont} ceux du sous-réseau mère. Dans chaque sous-réseau $N_j$, il faut au moins $deg(Q_j) - 1$ partages pour retrouver $Q_j(0)$.

Nous pouvons alors déterminer les différents seuils en fonction des degrés de nos polynômes. Tout d'abord, le nombre de réseaux $t_{networks}$ à compromettre dépend entièrement de $P$:

\begin{equation}
    t_{networks} = T(P)
\end{equation}

Avec $T$ le \og~seuil de déchiffrement d'un polynôme~\fg, c'est à dire son degré moins un. Le nombre de n\oe{}uds, tous réseaux confondus, nécessaires pour retrouver le secret est

\begin{equation}
\begin{split}
    &t_{nodes} = T(Q_0) + \\
    &min \left\{ \sum_{i \in I}T(Q_i) \bigg| I \subset \{1, \dots, l-1\}, |I| = T(P) - 1\right\}
\end{split}
\end{equation}

Enfin, un attaquant qui voudrait rendre le secret inaccessible devrait soit
\begin{itemize}
    \item éteindre $t_{f_0} = n_0 - T(Q_0) + 1$ n\oe{}uds dans le réseau mère, avec $n_0$ le nombre de n\oe{}uds dans le réseau mère~;
    \item éteindre $n_i - T(Q_i) + 1$ n\oe{}uds dans $x$ réseaux filles, avec $x = l - 1 - T(P) + 1$, et $n_i$ le nombre total de n\oe{}uds dans le sous-réseau $i$.
\end{itemize}

On a alors
\begin{equation}
\begin{split}
    &t_{f_1} = \\ 
    &min\left\{\sum_{i \in I} n_i - T(Q_i) + 1 \bigg| I \subset \{1, \dots, l-1\}, |I| = l - T(P)\right\}
\end{split}
\end{equation}

et donc

\begin{equation}
    t_{fail} = min\{t_{f_0}, t_{f_1}\}
\end{equation}

Le renouvellement des partages s'effectue comme pour le protocole LINCOS, tel que décrit en section~\ref{lincos_confidential}. Pour chaque sous-réseau, le propriétaire du document génère un nouveau polynôme $R_i \in \mathbb{F}_q[X], i \in [\![ 0, l-1 ]\!]$, tel que $R_i(0) = 0$, et distribue les évaluations de chaque polynôme $R_i$ aux différents n\oe{}uds du réseau $i$.

\section{Conclusion}

Dans cet article, nous proposons un nouveau protocole de stockage confidentiel à long terme sur plusieurs réseaux de distribution quantique de clé. Notre protocole, MULTISS, est une extension du protocole LINCOS, permettant de garantir la confidentialité à long terme y compris contre un adversaire en mesure de prendre le contrôle de tout un réseau de QKD.

En outre, notre protocole garantit la sécurité parfaite du document secret, c'est à dire que sa confidentialité ne pourra pas être altérée par l'évolution future de la puissance de calcul de l'attaquant.

\vspace{.5cm}

\subsubsection*{Remerciements}
T. Prevost remercie l'UCA pour son financement de thèse. Les auteurs sont reconnaissants à Bruno Martin (\url{https://webusers.i3s.unice.fr/~bmartin/}) pour son support théorique.

\bibliographystyle{IEEEtran}
\bibliography{main}
\vspace{12pt}

\end{document}